\documentclass[prd,aps,showpacs,
nofootinbib,
floatfix,superscriptaddress]{revtex4}
\usepackage{graphicx}
\usepackage{epsfig}
\usepackage{rotating}
\usepackage{amssymb}
\usepackage{dsfont}

\newcommand{\beq}{\begin{equation}}
\newcommand{\eeq}{\end{equation}}
\newcommand{\beqs}{\begin{eqnarray}}
\newcommand{\eeqs}{\end{eqnarray}}
\newcommand{\lsim}{\mathrel{\raisebox{-
.6ex}{$\stackrel{\textstyle<}{\sim}$}}}
\newcommand{\gsim}{\mathrel{\raisebox{-
.6ex}{$\stackrel{\textstyle>}{\sim}$}}}
\newcommand{\Tr}{{\rm Tr}}

\newcommand{\EFT}{effective field theory\,\,}
\newcommand{\VEV}{vacuum expectation value\,\,}
\newcommand{\EWSB}{electro-weak symmetry breaking\,\,}
\newcommand{\SM}{standard model\,\,}
\newcommand{\CFT}{conformal field theory\,\,}

\def\hbar{\hspace{0pt}\raisebox{1pt}{$-$} \hspace{-7pt} h}

\def\di{\mbox{d}}

\begin{document}
\title{Dynamical electro-weak symmetry breaking from deformed AdS:\\
vector mesons and effective couplings.}

\author{Marco Fabbrichesi}
\affiliation{INFN, Sezione di Trieste, Trieste, Italy.}
\author{Maurizio Piai }
\affiliation{Department of Physics, University of Washington, Seattle, WA 98195}
\affiliation{Swansea University, School of Physical Sciences,
Singleton Park, Swansea, Wales, UK}
\author{Luca Vecchi}
\affiliation{INFN, Sezione di Trieste, Trieste, Italy.} 
\affiliation{Scuola Internazionale Superiore
di Studi Avanzati, Via Beirut 4, Trieste, Italy.}
\date{\today}

\begin{abstract}
We study a modification of the five-dimensional  description of
dynamical electro-weak symmetry breaking inspired by the AdS/CFT
correspondence. Conformal symmetry is broken in the low-energy region near the IR brane by a power-law departure from the pure AdS background. 
Such a modification---while not
spoiling the identification of the IR brane with the scale of
confinement--- has a dramatic effect on both the coupling of
the first composite states to the standard model currents and
their self-couplings. Chiral symmetry breaking can take place at
a scale larger than the IR cut-off. This
study shows that  observables, such as the precision
parameter $\hat{S}$, which depend on the couplings of the lightest
composite states to the currents are very sensitive to the details
of the dynamics in the low energy region where conformal symmetry
is lost and  electro-weak symmetry is broken just above the scale
of confinement. Therefore results of calculations
 of these observables in AdS/CFT inspired scenarios
should be interpreted conservatively.
The most important phenomenological consequence for physics at the LHC is
that the  bound on the mass scale of the heavy excitations
(technirho mesons) in a realistic model is in general lower than in the
pure AdS background with a simple hard-wall cut-off in the IR.

\end{abstract}


\maketitle

\section{Introduction}

The mechanism responsible for \EWSB in the \SM will be tested at
the Large Hadron Collider (LHC). One important goal for this
experimental program is to understand whether the  interactions
responsible for \EWSB are strong or weak. It is essential to
identify theoretically clean, measurable quantities that can help
distinguish these two possibilities unambiguously.

One might think that this is an easy task: after all, a strongly
coupled model in the spirit of technicolor~\cite{TC} predicts the
existence of towers of broad, strongly coupled composite
resonances, with a rich spectroscopy at the TeV scale, in analogy
with what is known about QCD at the GeV scale. Yet, indirect
experimental data about \EWSB\cite{PT,Barbieri} cannot be
easily reconciled with this framework and already suggest that, if
\EWSB is a dynamical effect, the low-energy \EFT description of
the new strongly-coupled sector has to exhibit features that are
not generic. Walking technicolor~\cite{walking} is a plausible
candidate for such a strongly coupled model, based on conformal
behavior and large anomalous dimensions in the IR, but
calculability within this framework has been a
challenge~\cite{AS}.

In recent years, based on the ideas of the AdS/CFT
correspondence~\cite{AdS/CFT}, and on the pioneering work of
Randall and Sundrum~\cite{RS1,pheno}, many models have been
investigated which exhibit in the low energy region the basic
properties expected in a walking theory, while being calculable.
Examples now exist of models that are compatible (within the
errors) with the precision data and can be discovered at the LHC.
The literature on the subject is already extensive~\cite{AdS/TC,HS,MP,higgsless,composite}.
Most of these models assume a conformal behavior of the strongly coupled
sector in the energy region spanning few orders of magnitude above
the electro-weak scale and the existence of a weakly-coupled \EFT
description of the low-energy dynamics of the resonances. The
construction of  the \EFT is derived by writing a weakly coupled
extra-dimension model with a non-trivial gravity background, and
by using the dictionary of the AdS/CFT correspondence to relate
back to four dimensions. A generic phenomenological feature of all
these models is that, unless a clever mechanism arranging for
non-trivial (often fine-tuned) cancellations is implemented, a
quite severe lower bound on the mass $M_1 \gsim$ 2.5--3 TeV of
the lightest  spin-1 resonance (techni-rho) results, in particular
from the bounds on the electro-weak parameter $\hat{S}$~\cite{PT,Barbieri}. This result,
together with the assumption that the \EFT be weakly coupled (and
hence calculable), gives rise to  a spectacular signature (a sharp resonance peak) at
the LHC~\cite{MP2}. Unfortunately, it is very difficult to distinguish it from
the signature of a generic, weakly-coupled extension of the \SM
with an extended  gauge group, predicting a new massive
$Z^{\prime}$ gauge boson.

Indisputable evidence proving that a  strongly-coupled sector is
responsible for \EWSB would be the discovery of at least the first
two spin-1 resonances, hence proving that these new particles are
not  elementary, but higher energy excitations of a composite
object. The major obstacle against this scenario is the
unfortunate numerology emerging from the combination of precision
data and LHC high-energy discovery reach. If $M_1 \gsim$ 2.5--3 
TeV, than it follows that the mass of the second resonance must be $M_2
\gsim$ 5--6 TeV and just beyond the region where LHC data are
expected  to give convincing evidence~\cite{AtlasTDR}. Yet, a
pretty mild relaxation of the experimental bounds would be enough
to change this situation radically, since $M_1 \sim 1.5$ TeV would
imply $M_2 \approx$ 2.5--4 TeV, well within reach even at
moderate luminosity~\cite{MP2}. It is hence timely, just before
LHC starts collecting data, to question how accurate the AdS/CFT
description of  realistic dynamical \EWSB is, and whether some of
the approximations implied by this description could account for
the desired softening of the bounds, without at the same time
spoiling the calculability of the effective field theory.

In analogy with~\cite{RS1}, the five-dimensional picture usually contains
two hard boundaries representing the UV and IR cut-off between
which the theory is conformal. This is the weakest link with the
idea that \EWSB be triggered by a non-abelian gauge theory with an
approximate IR fixed point. Taken literally, this picture means
that, both in the UV and in the IR, conformal symmetry is lost
instantaneously, via a sharp transition. As for the UV cut-off,
this is not a real problem from the low-energy \EFT point of view.
The details of how an asymptotically-free fundamental theory in
the far UV enters a quasi-conformal phase below the UV cut-off,
can always be reabsorbed (via holographic
renormalization~\cite{HR,MP}) in the definition of otherwise
divergent low-energy parameters of the effective field theory,
defined at a given order in the perturbative expansion of the \EFT
itself.

Rather different is the case against using a hard-wall regulator
in the IR. There is no sense in which IR effects decouple and can
be renormalized away, and hence  the low-energy effects we are
interested in, when comparing the \EFT to the experimental data,
are inherently sensitive to the choice of the IR regulator. On the
one hand, the very validity of the \EFT description based on the
AdS/CFT dictionary requires that the hard-wall cut-off be at least
a reasonable leading order approximation (otherwise the \EFT
itself would be strongly coupled, and not admit a controllable
expansion). On the other hand, corrections are expected to be
present, and  estimating their size and understanding their
phenomenological consequences is crucial, at the very least in
order to know what to expect in  experiments such as those at the LHC,
which is going to test precisely the energy range close to the IR
cut-off.

To be more specific.
In the IR, three different phase transitions are taking place:
electro-weak symmetry breaking, conformal symmetry breaking
and confinement. These cannot define three parametrically separate scales,
since they are all triggered by the same physical effect, namely the fact that
the underlying (unknown) theory possesses an approximate fixed point in the IR.
Hence the RG flow of the underlying dynamics
is not going to reach the IR fixed point (which is only approximate),
but will drift away from it  at low energies, after spending some time
(walking) in its proximity.
Yet, there is no reason to expect these three effects
to arise precisely at the same energy (temperature), and
they might define three distinct critical scales (temperatures)
that differ by $O(1)$ coefficients.

An illustration of this point can be obtained by considering an
${\cal N}=1$ supersymmetric QCD model with $N_c$ colors and $N_f$
fermions.
  At least at large-$N_c$, for $3N_c/2<N_f<3N_c$, the theory is asymptotically free,
   but  has a fixed point in the IR~\cite{seiberg,cascade} (for recent progress towards the rigorous
   construction of the gravity dual see, for instance, \cite{Nunez}). If $N_f$ is not far from the lower
    bound,
   so that the theory is strongly coupled at distances larger than
   a UV cut-off $1/L_0$, then the theory might be approximately described by
   a large-$N_c$ \CFT at strong 't Hooft coupling.
   Suppose now that at some smaller energy,
   characterized by a length scale $\bar{L}\gg L_0$,
    for some reason (for example the existence of a suppressed
   symmetry-breaking higher-order operator, which acquires a large anomalous dimension
   in the IR turning it into a relevant deformation)
   a symmetry-breaking condensate forms,
   reducing further $N_f$ to a value $N_f^{\prime}$ closer to or below $3N_c/2$.
   Symmetry-breaking drives the theory away from the original fixed point,
   and induces the loss of conformal symmetry.
   The coupling now runs fast (because the coupling itself was already big
    and large anomalous dimensions are present),
   and (depending on $N_f^{\prime}$)
   the theory either  enters a new conformal phase at stronger coupling
   or  confines.
The breaking of the global $SU(N_f)_L\times SU(N_f)_R$,
conformal symmetry-breaking and confinement
take all place approximately at the same  scale. Yet, the energy at which the coupling
reaches its upper bound defines a new scale $L_1$
which might well be some numerical factor  away from $\bar{L}$, the scale
at which the RG-flow trajectory departed away from the fixed point.

If this is the qualitative behavior of the UV-complete dynamical
model that is ultimately responsible for \EWSB, describing it as a
slice of AdS space between two hard walls is a good leading-order
approximation. Nevertheless, we may wonder whether a  factor of 3 or 4
separating the scales of conformal symmetry-breaking and
confinement can be completely ignored, in the light of the
phenomenological consequences at the LHC that a mere factor of two
might have. In this paper, we study the effect of such a factor. We
consider the simplest possible \EFT description of
dynamical \EWSB  as a 5D weakly-coupled system (see
also~\cite{MP}), introduce (besides the UV brane at $L_0$ and the
IR brane at $L_1$) a new discontinuity at the scale $\bar{L}$,
very close to the IR scale $L_1$, and assume that the background
deviates from the AdS case for $\bar{L}<z<L_1$.

As for the origin and description of electro-weak symmetry-breaking,
we will treat  it as a completely non-dynamical effect
localized in the IR, somehow in the spirit of Higgless models.
The breaking could take place at $L_1$ as well as at $\bar{L}$ (or anywhere in between),
as suggested by the SQCD example above.
We compare the effects on the electro-weak
precision parameter $\hat{S}$ in these two cases,
as illustrative of two extreme possibilities,
without committing ourselves to either of them.
The idea that chiral symmetry breaking might, for a generic
model, take place at a scale higher than confinement
has been in the literature for a while~\cite{chiral/confinement},
has been supported by lattice evidence in some special case~\cite{lattice},
and has recently been discussed also in string-inspired models~\cite{SS}.

A realistic model should also implement a dynamical mechanism
generating the mass of the \SM fermions. This can be done either
via extended technicolor higher-order interactions between the \SM
fermions and the new strong sector~\cite{ETC,APS}
(represented in the 5D picture  by Yukawa interactions localized
at the UV, with the symmetry-breaking \VEV not localized, but
exhibiting a non-trivial power-law profile in the bulk), or via
the assumption that \SM fields are themselves (partially)
composite, in the spirit of topcolor and
related models~\cite{topcolor} (which would imply the fermions be allowed
to propagate in the bulk of the 5D model). A detailed discussion
of  how the global  family symmetry of the \SM is  broken would be
required in order to study how the phenomenology of
flavor-changing transitions and the physics of the third
generation would be affected by the proposed modification of the
background. In this paper,
we treat the \SM fermions as non-dynamical fields, described by a
set of external currents, and do not address the problem of their
mass generation.

\section{Preliminaries}

A non-trivial departure of the dynamics of the spin-1 resonances,
with respect to that on pure AdS geometry, may be either due to a
modification of the gravity background or to the presence of a
non-dynamical background (dilaton). Since we consider
an \EFT where only spin-1 states are dynamical, it is not possible
to distinguish between these two effects at this level. We choose
to describe the model in terms of a deformation of the gravity
background, for simplicity.

Consider the five-dimensional space described by the metric
\beqs
\label{Eq:AdS} \di s^2&=&a(z)^2\left(\eta_{\mu\nu}\di x^{\mu}\di
x^{\nu} \,-\,\di z^2\right)\,, \eeqs
where $L_0<z<L_1$. We will
assume that the geometry approaches pure AdS in the UV region,
$a(z)\rightarrow L/z$ as $z\rightarrow L_0$, and departs from it
at a scale $z\sim\bar L$. In most of the calculations we take
$L_0=L$ for simplicity.

We are interested in describing a model that at  low-energy (below
$1/L_1$) can be matched to the electro-weak chiral
Lagrangian~\cite{EWCL}. This requires to introduce a 5-dimensional
gauge group which is at least $SU(2)_L\times U(1)_Y$, but may be
enlarged to accommodate custodial symmetry. Irrespectively of the
details, the model contains a vectorial sector (the neutral part
of which consists of the photon and its excitations) and an axial
sector (containing the $Z$ boson and its excitations). In this
paper we  describe only the phenomenology connected
with the neutral gauge bosons, hence we dispense with the details
of the complete symmetry group. For concreteness, we take the
vectorial sector to be described by the pure Yang-Mills $SU(2)$
theory with the following action:
\beqs {\cal S} &=&
\int\di^4x\int_{L_0}^{L_1}\di z\sqrt{G}
\left[G^{MR}G^{NS}\left(-\frac{1}{2}\Tr F_{MN}F_{RS}\right)+2g
\delta(z-L_0)G^{MN}\Tr J_MA_N\right]\,, \eeqs
where
$F_{MN}=\partial_MA_N-\partial_NA_M+ig[A_M,A_N]$ is the field
strength tensor of $A_M=A_M^aT^a$ with $T^a=\tau^a/2$ the
generators of $SU(2)$, where $g$ is the (dimensionful) gauge
coupling, and where $J_M=(J_{\mu}(x),0)$ is the four-dimensional
external current localized on the UV-brane.

Quantization requires to add appropriate gauge-fixing terms,
canceling the mixing terms between spin-0 and spin-1 fields,
which in unitary gauge implies $A_5(x,z)=0$.

After Fourier transforming in 4D,
$A_{\mu}(x,z)\equiv\int\frac{\di^4 q}{(2\pi)^2}
e^{iqx}A_{\mu}(q)v(q,z)$, the free bulk equations read \beqs
\label{Eq:bulk1}
\partial_5\left[a(z)\partial_5 v(q,z)\right]&=&-q^2a(z) v(q,z)\,.
\eeqs
Substituting the solutions in the action, and canceling the boundary terms
at $z=L_1$, without breaking the
gauge symmetry, requires to impose Neumann boundary conditions:
\beqs
\label{Eq:Neumann}
\partial_5 v(q,L_1)&=&0\,.
\eeqs
This set of equations admits always a constant, massless zero mode.

Finally, the action can be rewritten as a pure boundary term at
the UV, from which one can read the vector two-point  correlator,
that for $L_0\rightarrow L$ is
\beqs
\Sigma_V(q^2)&=&
g^2\frac{v(q,L_0)}{\partial_5v(q,L_0)}\,, \label{Eq:correlator}
\eeqs which can be expanded as 
\beqs
\Sigma_V(q^2)&=&e^2\left(\frac{1}{q^2}\,+\,\sum_i\frac{R_i}{q^2-M_i^2}\right)\,,
\eeqs
where $M_i$ ($i=1,2,\dots$) are the masses of the excited
states, and $R_i$ define their effective couplings to the
four-dimensional currents, normalized to the coupling $e^2$ of the
massless mode (to be identified with the electro-magnetic coupling
of the photon).

The (dimensionful) bulk coupling $g$ controls the perturbative
expansion used to extract this correlator. It is not directly
related to the effective coupling $e$ of the \SM gauge boson
(photon), but rather is related to the strength of the effective
interactions among its heavy (composite) excitations. The
relation between these two effective couplings depends on how the
theory is regularized in the UV, and is not a calculable quantity,
because of the divergences in the $L_0\rightarrow 0$ limit. A
rigorous treatment requires to introduce appropriate counterterms
and treat the ratio $e^2L/g^2$ as a free parameter. For the
purposes of this paper, which primarily require comparing
identical UV settings with different IR deformations, we can
simplify this procedure by assuming that $L_0\ll L_1$ be finite
and fixed, and express this ratio as a function of the scales and
couplings in the model. We  discuss later  how
good the perturbative expansion is by estimating the size of the
effective self-coupling of the composite states.

In order to compute $\hat{S}$ one has to introduce also the
axial-vector excitations, and a symmetry-breaking mechanism. For
the purposes of this paper, we only consider the Higgsless limit,
defined by the introduction of a localized, infinitely massive
Higgs scalar which assumes a non-trivial symmetry-breaking vacuum
expectation value.

The axial-vector modes $v_A(q,z)$ still satisfy
Eq~(\ref{Eq:bulk1}), but their boundary conditions  (and the gauge
fixing action)  are modified. We consider two cases in the
following. In the first, symmetry-breaking takes place on the
boundary $L_1$ so that the axial-vector profiles $v_A(q,z)$ obey
generalized Neumann boundary condition: \beqs
\partial_5v_A(q,L_1)\,+\,m v_A(q,L_1)&=&0\,.
\eeqs The effective symmetry-breaking parameter $m$ has dimension
of a mass. In the limit $m\rightarrow 0$ one recovers the
symmetric case, while for $m\rightarrow +\infty$ one recovers the
Dirichelet boundary conditions. The mass of the $Z$ boson depends
on $m$ is such a way that it vanishes for vanishing $m$, but is
determined by $L_1$ for arbitrarily large $m$. In the second case
we consider a symmetry-breaking \VEV localized at a different
point $\bar{L}<L_1$ in the fifth dimension. The modifications to
be implemented in this case will be discussed in the next
sections.

All of this allows to define the axial-vector correlator $\Sigma_A(q^2)$ by replacing in Eq.~(\ref{Eq:correlator})
$v_A(q,z)$ and its derivative to $v(q,z)$.
After these manipulations, the precision parameter $\hat{S}$ is given by
\beqs
\hat{S} &=& \left.e^2\cos^2\theta_W \,\frac{\di}{\di q^2}\left(\frac{1}{\Sigma_V(q^2)}-\frac{1}{\Sigma_A(q^2)}\right)\right|_{q^2=0}\,,
\eeqs
where $e$ has been defined before, and corresponds to the electro-magnetic coupling,
while  $\theta_W$ is the effective weak-mixing angle.
We recall here that an approximate extrapolation
to large Higgs masses yields the experimental
limit $\hat{S}\lsim 0.003$ at the $3\sigma$ level~\cite{Barbieri}.

\section{Pure AdS background}

We summarize here the results of the case in which the background
is purely AdS with $a(z)=L/z$, and assume for simplicity that
$L_0=L$. The vector correlator is \beqs \Sigma^{(0)}_{V}(q^2)&=&
\frac{g^2 \left(J_0(L_1 q) Y_1(L q)-J_1(L q)
   Y_0(L_1 q)\right)}{q \left(J_0(L_1 q) Y_0(L
   q)-J_0(L q) Y_0(L_1 q)\right)}\,.
   \eeqs
In order to discuss the spectrum and couplings, the following approximations
can be used:
\beqs\label{app1}
\Sigma^{(0)}_{V}(q^2)&\simeq&
\frac{g^2 J_0(L_1q)}
{L q^2 \left(\frac{\pi}{2}Y_0(L_1 q)-J_0(q L_1)\left(\gamma_E+\log\frac{L q}{2}\right)\right)}\\\label{app2}
&\simeq&
\frac{g^2}
{L q^2 \left(\frac{\pi}{2}\tan(L_1 q - \frac{\pi}{4})-\left(\gamma_E+\log\frac{L q}{2}\right)\right)}\,,
\eeqs
the first of which is valid for $L\ll L_1$, and the second for $qL_1 >1$.

From~(\ref{app1}) one can read the coupling of the zero mode:
\beqs\label{ee}
e^2&=&\frac{g^2}{L\log L_1/L}\,.
\eeqs
From~(\ref{app2}) one can look for the poles and the residues $R_i$.
The poles (besides the pole at zero)
are in the vicinity of those  of $\tan(L_1 q-\pi/4)$:
 \beqs
 M_i&\simeq&\frac{\pi}{4L_1}\left((4i-1)-\frac{2}{\gamma_E+\log (4i-1)\pi L/(8L_1)}\right)\,,
 \eeqs
while the residues are approximately given by:
\beqs R_i&\simeq&
\frac{4\log(L_1/L)} {-2+\frac{2\pi L_1M_i}{1+\sin(2L_1M_i)}},
\eeqs
with $i=1,2,\dots$. These approximations are acceptably
accurate as long as $L\ll L_1$. A numerical calculation will be
performed later on, when discussing the phenomenology for some
relevant choice of parameters.

The axial correlator can be computed exactly:
\beqs
 \Sigma^{(0)}_{A}(q^2)&=&
\frac{g^2 \left(\left(q J_0(L_1 q)+m
   J_1(L_1 q)\right) Y_1(L q)-J_1(L q) \left(q
   Y_0(L_1 q)+m Y_1(L_1
   q)\right)\right)}{q \left(q J_0(L_1 q) Y_0(L
   q)+m J_1(L_1 q) Y_0(L q)-J_0(L q) \left(q
   Y_0(L_1 q)+m Y_1(L_1
   q)\right)\right)}\,,
\eeqs
and, for $L_0\ll L_1$, yields
\beqs
\hat{S}&=&\frac{\cos^2\theta_W L_1 m (3 L_1
   m+8)}{4 (L_1 m+2)^2 \log
   \left(\frac{L_1}{L}\right)}\,.
\eeqs

In the limit $m\rightarrow +\infty$ we have
\beqs
\hat{S}&=&\frac{3\cos^2\theta_W}{4\log{L_1/L}}\,.
\eeqs
Imposing
the ($3\sigma$-level) experimental limit we find that
\beqs
\frac{g^2}{L}&=&e^2\log\frac{L_1}{L}\,=\,e^2\frac{3\cos^2\theta_W}{4\hat{S}}\,\gsim\,20\,,
\eeqs
where $e^2\simeq 0.1$ is the effective coupling of the
electro-magnetic $U(1)_Q$ in the standard model. Since, as discussed later, $g^2/L$ gives a measure of the effective
strength of the self interactions between resonances (and the
dimensionful coupling $g$ is the expansion parameter in the 5D
action) the experimental bounds are satisfied only at the price of
loosing calculability, as is the unfortunate case also when trying
to build QCD-like technicolor models in 4D, either using the
large-$N$ expansion, hidden local symmetry, or deconstruction (see
for instance~\cite{deconstructTC}). We do not discuss further
this limit.

In the more interesting and realistic case in which $m L_1 \ll 1$,
the axial-vector spectrum and couplings are approximately the same
as the vectorial sector. In this framework $m$ is just a free
parameter, and we treat it as such. With finite $mL_1 \ll1$, the
mass of the lightest axial-vector state is approximately
$M_Z^2\simeq m/(L_1\log (L_1/L))$, and hence 
\beqs
\hat{S}\simeq\frac{\cos^2\theta_W}{2\log L_1/L}\,mL_1
\simeq\frac{\cos^2\theta_W}{2}M_Z^2L_1^2\,
\eeqs
satisfies the bounds on $\hat{S}$ for $1/L_1\gsim 1$ TeV, which
depending on the value of $L_0/L_1$ translates into a bound
$M_1\gsim 2.5 - 4$ TeV. For instance, for $g^2/L<1/2$
it requires $M_1 \gsim 2.8$ TeV, and consequently $M_2\gsim 6$ TeV,
which is beyond the projected reach of the LHC searches.

\section{Departure from AdS}

We now consider the possibility that conformal invariance be violated at
some energy regime above the confinement scale and suppose there
exists a hierarchy of scales $L_0=L<\bar{L}<L_1$ such that the
space is the usual AdS for $L_0<z<\bar{L}$, but departs from it in
the IR region $\bar{L}<z<L_1$. Our aim is to model this behavior
without affecting the approximate description of confinement
provided by the IR brane (different motivations lead the authors
of~\cite{HS} to other parameterizations). The simplest form one can
choose in order to achieve this goal is a power-law warp factor
\beqs\label{Eq:Bckgrnd} a(z)&=&\left\{\begin{array}{cc}
\frac{L}{z}&z<\bar{L}\,\cr
\frac{L}{z}\left(\frac{\bar{L}}{z}\right)^{n-1}&z>\bar{L}
\end{array}\right.\,.
\eeqs 

This parameterization may be viewed as a leading order
approximation of a smooth background describing the appearance of
some relevant deformation in the \CFT before the underlying
fundamental theory confines.

We will see later that a power-law avoids generating an explicit
mass gap from the bulk equations, so that the quantity $1/L_1$ can
still be interpreted as the scale of confinement. Moreover, with
our parameterization we can solve the equations exactly and in a
very straightforward way, which is in itself a welcome property
when modeling an otherwise untreatable dynamical system.

Most of the algebraic manipulations  can be performed for
generic $n$. Yet,  we discuss explicitly only the $n>1$ case.
 A variety of arguments, all ultimately descending from unitarity,
suggest that we should limit ourselves to $n\geq 1$. An extra-dimensional argument
 can be derived along the lines of~\cite{weakenergy}, in which it is
 shown how the weaker energy condition leads to a $c$-theorem controlling
the behavior of the  curvature in crossing a phase transition
towards the IR. This is related to the fact that, in the context
of strongly-coupled four-dimensional models, in going through a
phase transition it is reasonable to expect the effective number
of light degrees of freedom to decrease~\cite{Appelquist}. Hence the effective coupling of the \EFT description, which is
related to the $1/N$ expansion, is expected to increase. We 
show later in the paper  that the effective self-coupling
 of the heavy resonances is enhanced for $n\geq 1$, in agreement with
 the four-dimensional intuitive expectation, and that this enhancement is controlled by a power
 of the ratio of relevant scales, in agreement with naive expectations for
 a theory with a generic deformation due to a relevant operator.
The fact that all of our results agree with the intuitive
interpretation gives an indication in support both of  the
power-law parameterization chosen here and of the $n\geq 1$
restriction.

The solutions to the bulk equations in the IR region $z>\bar{L}$ are of the form
\beqs
v^{IR}(q,z)&=&z^{\frac{n+1}{2}}\left(c_1^{IR}(q)J_{\frac{n+1}{2}}(qz)\,
+\,c_2^{IR}(q)Y_{\frac{n+1}{2}}(qz)\right)\,,
\eeqs
while in the UV region
\beqs
v^{UV}(q,z)
&=&z\left(c_1^{UV}(q)J_1(qz)\,+\,c_2^{UV}(q)Y_1(qz)\right)\,.
\eeqs
The bulk profile is obtained by applying the IR boundary conditions to
$v^{IR}$, and then by requiring that the junction of the two solutions
be smooth, so that no boundary action localized at $\bar{L}$ is left:
\beqs
\partial_5v^{IR}(q,L_1)&=&0\,,\\
v^{IR}(q,\bar{L})&=&v^{UV}(q,\bar{L})\,,\\
\partial_{5}v^{IR}(q,\bar{L})&=&\partial_{5}v^{UV}(q,\bar{L})\,.
\eeqs
The correlator is then obtained from  Eq.~(\ref{Eq:correlator}) by using $v^{UV}$.
From all of this, one can extract the masses and couplings of the resonances.
In particular, the coupling of the zero-mode (photon) is
\beqs
e^2&=&\frac{(n-1)\frac{g^2}{L}}{(n-1)\log(\frac{\bar
L}{L_0})+\left(1-\left(\frac{\bar L}{L_1}\right)^{n-1}\right)}\,.
\eeqs
For $n=1$, or for $\bar{L}=L_1$,
one recovers the AdS result~(\ref{ee}). For $n>1$ and $\bar{L}<L_1$ this estimate is
enhanced (for fixed $g^2/L$). In order to understand how significant this
effect is, one needs to compare this coupling to the effective self-coupling,
which is discussed in the next section.

Analytical expressions for the couplings and masses of the
vector-like resonances are rather involved, and numerical results
will be plotted later on. In order to gain a semi-quantitative
understanding of how these quantities are modified with respect to
the pure AdS case, we discuss the (unrealistic) extreme case in
which $\bar{L}=L_0 \ll L_1$. For $qz\gg1$: \beqs
J_{\frac{n-1}{2}}(qz)&\simeq&\sqrt{\frac{2}{\pi qz}}\cos\left(qz-\frac{n\pi}{4}\right)\,,\\
Y_{\frac{n-1}{2}}(qz)&\simeq&\sqrt{\frac{2}{\pi
qz}}\sin\left(qz-\frac{n\pi}{4}\right)\, \eeqs and the masses of
$i$-th resonances, for $n\gsim 1$, are approximately given by the
zeros of $J_{\frac{n-1}{2}}(qL_1)$, \beqs
M_i(n)&\simeq&\frac{2i-1}{2}\frac{\pi}{L_1}\,+\,\frac{n\pi}{4L_1}\,,
\eeqs with $i=1,2,\dots$. This agrees with the pure AdS case
($n=1$) at least for $L/L_1\ll 1$. For the more realistic case in
which $L_0 \ll \bar{L} < L_1$, the spectrum of heavy modes is
going to be shifted, with masses heavier  by approximately
$(n-1)\pi/(4L_1)$ with respect to the AdS case, for those resonances
whose masses are comparable with the new scale $1/\bar{L}$. The
spectrum connects back to the pure AdS case for higher excitation
number $i$.

As for the residues, the couplings to the
currents of the heavy modes are approximately going to be suppressed
with a power-law dependence
$\approx \left(\frac{\bar{L}}{L_1}\right)^{(n-1)}$
with respect to the AdS case. Again, this suppression applies only to
the lightest resonances,  those for which   the mass is shifted to higher values.

For the axial-vector case, if the symmetry breaking takes place at $L_1$
the only change is the IR boundary condition:
\beqs
\partial_5v_A^{IR}(q,L_1)\,+\,mv_A(q,L_1)&=&0\,,\\
v_A^{IR}(q,\bar{L})&=&v_A^{UV}(q,\bar{L})\,,\\
\partial_{5}v_A^{IR}(q,\bar{L})&=&\partial_{5}v_A^{UV}(q,\bar{L})\,.
\eeqs For generic $n>1$, we find the following approximations for
the mass of $Z$ boson and for $\hat{S}$: \beqs
M_Z^2&\simeq&\left(\frac{\bar{L}}{L_1}\right)^{n-1}\frac{(n-1)m}
{L_1\left(1-(\bar{L}/{L_1})^{n-1}+(n-1)\log \bar{L}/L\right)}\,,\\
\hat{S}&\simeq&
\cos^2\theta_W\left(\frac{1}{n+1}+\frac{1}{2}(\bar{L}/L_1)^2-\frac{1}{n+1}(\bar{L}/L_1)^{n+1}\right)L_1^2M_Z^2\,,
\label{Eq:ShatL1} \eeqs where the last approximation is valid as
long as $mL_1\ll1$. For small  $\bar{L}/L_1$ this approximation
would not hold, because of the dependence of $M_Z$ on $m$ and on
$\bar{L}/L_1$. We do not admit a parametric separation between
$\bar{L}$ and $L_1$, and hence the approximations are acceptable.
We also checked this numerically, using the exact bulk profiles
and correlators.

The other extreme  possibility we are interested in
 is the one in which the symmetry-breaking condensate is localized at $\bar{L}$,
for which the boundary conditions become
\beqs
\partial_5v_A^{IR}(q,L_1)&=&0\,,\\
v_A^{IR}(q,\bar{L})&=&v_A^{UV}(q,\bar{L})\,,\\
\partial_{5}v_A^{IR}(q,\bar{L})&=&\partial_{5}v_A^{UV}(q,\bar{L})\,+\,\bar{m}v_A(q,\bar{L})\,.
\eeqs
For generic $n$:
\beqs
M_Z^2&\simeq&\frac{(n-1)\bar{m}}{\bar{L}\left(1-(\bar{L}/L_1)^{n-1}+(n-1)\log \bar{L}/L\right)}\,,\\
\hat{S}&\simeq&\frac{\cos^2\theta_W\left(n+1-2\left(\bar{L}/L_1\right)^{n-1}\right)}{2(n-1)}\bar{L}^2M_Z^2\,
\label{Eq:ShatLb}
\eeqs
in which in the last expression only the leading order of the expansion in $M_Z$
has been kept, and all the expressions are valid
as long as $\bar{m}L_1 \ll 1$.

Notice how the dependence of $M_Z$ on $\bar{m}$ 
is not suppressed by powers of $\bar{L}/L_1$, as in the former
case, where $m$ came from a localized term at $L_1$. This result
agrees with the intuitive notion that moving the symmetry-breaking
towards the UV enhances its effect for the zero-mode, while
suppressing the mass splitting of the heavy resonances. The result is well illustrated by $\hat{S}$, which is proportional to
$M_Z^2$ through the position $\bar{L}$ or $L_1$ of the
symmetry-breaking condensate in the fifth dimension.

\section{Estimating the strength of the self-interactions}

The departure from conformal invariance,  explicitly added via a power-law
deviation from the AdS background in the IR region, might imply
that the dynamics of the \EFT itself be strongly coupled, as is
the case for a QCD-like dynamical model. It has to be understood
if the \EFT treatment still admits a power-counting allowing to
use a cut-off $L_0$ much larger than the electro-weak
scale. A fully rigorous treatment of this problem is not possible,
because it requires to extend the \EFT Lagrangian beyond the
leading order in $1/N_c$. Yet, a reasonable estimate of the
effective coupling can be extracted by looking at the cubic and
quartic self-couplings of the resonances, the structure of which
(at the  leading order) is dictated by 5D gauge-invariance.

Consider first the pure AdS background and define \beqs
g_{\rho}^{(i)\,2}&\equiv&\frac{g^2}{L}\frac{\int_{L_0}^{L_1}\frac{\di
z}{z}|v(M_i,z)|^4} {\left(\int_{L_0}^{L_1}\frac{\di
z}{z}|v(M_i,z)|^2\right)^2}\,. \eeqs The expansion parameter is
related to $g_{\rho}$, which we define as the asymptotic limit of
the effective self-coupling for large excitation number. As long
as $L_0 \ll L_1$ and $M_iL_1\gg1$, the bulk profiles of the heavy
modes can be approximated by 
\beqs v(M_i,z)
&\propto&\frac{z}{\sqrt{M_i}}\,J_1(M_i
z)\,\propto\,\frac{\sqrt{z}}{M_i}\cos\left(M_iz-\frac{3\pi}{4}\right)\,
\eeqs 
yielding 
\beqs g_{\rho}^2\,\equiv\,\lim_{i\rightarrow
+\infty}g_{\rho}^{(i)\,2}&\simeq&\frac{3}{4}\frac{g^2}{L}\,. 
\eeqs
For the smallest values of $i=1,2$ this is a moderate
underestimate. For instance for $i=1$, from the exact solution one
obtains $g_{\rho}^{(1)\,2}\sim 1.2 g^2/L$. The meaning of this
definition of $g_{\rho}$ is that it gives a reasonable estimate
of the strength of the self-coupling of the resonances, and hence
of the expansion parameter of the \EFT (which is related to the
large-$N_c$ expansion). As expected, this turns out to be
controlled by $g^2/L$, up to $O(1)$ coefficients. The actual value
of $g^2$ is related (with the treatment of the UV cut-off used
here) to the coupling of the zero mode $e^2=g^2/(L \log L_1/L_0)$,
so that  $g_{\rho}^2 \approx  e^2 \log (L_1/L_0)$. This yields the
relation between strength of the effective coupling and the
effective cut-off in the UV, which as expected is logarithmic,
ultimately because of conformal symmetry. The requirement that
this defines a perturbative coupling $g_{\rho}^2$ implies
a bound on $L_1/L_0$. Choosing for instance $L_1=100 L_0$ (a
value that is not justifiable by applying naive dimensional analysis to the electro-weak
chiral Lagrangian), yields $g_{\rho}^{(i)\,2} \approx 0.3$, which
means that the \EFT admits an acceptable expansion in powers of
$g_{\rho}^2/(4\pi)$ even with large choices of the UV
cut-off $1/L_0$.

Generalizing this estimate in presence of the non-trivial
background~(\ref{Eq:Bckgrnd}) is somehow more difficult, largely
because of the junction conditions at $\bar{L}$. This can be done
numerically, but for the present purposes a semi-quantitative
assessment of the size of the effective coupling suffices. We
again focus on large values of $M_iL_1$ and modify the definition
of the effective couplings to \beqs
g_{\rho}^{(i)\,2}&\equiv&\frac{g^2}{L}\frac{\int_{L_0}^{L_1}\frac{\di
z}{z^n}|v(M_i,z)|^4} {\bar{L}^{n-1}\left(\int_{L_0}^{L_1}\frac{\di
z}{z^n}|v(M_i,z)|^2\right)^2}\,. \eeqs

The specific case we are interested in lies somewhere in between
the pure AdS and the pure power-law. In the latter case an
acceptable approximation would be: \beqs v(M_i,z)
&\propto&\frac{z^{\frac{n+1}{2}}}{\sqrt{M_i}}\,J_{\frac{n+1}{2}}(M_i
z)\,\propto\,\frac{z^{\frac{n}{2}}}{M_i}\cos\left(M_iz-\frac{(n+2)\pi}{4}\right)\,.
\eeqs The effective coupling receives power-law contributions in
$L_1/\bar L$, plus terms that are logarithmic in $L_0/\bar{L}$ and
hence subleading $O(1)$ corrections. The power-law is the most
important effect and, for largish choices of $L_1/\bar{L}$ and in
the case $n>1$, we obtain: 
\beqs
g_{\rho}^{2}&\simeq&\frac{3}{2(n+1)}\frac{g^2}{L}\left(\frac{L_1}{\bar{L}}\right)^{n-1}\,
\label{Eq:grho}\\
&\simeq&\frac{3e^2}{2(n^2-1)}\left(\frac{L_1}{\bar{L}}\right)^{n-1}\,
\eeqs 
which, as in the pure AdS case, represents a defective
approximation by roughly a factor of 2 for the very first
resonance. We see that, for $g_{\rho}$ to be acceptably small as
to define an expansion parameter, $L_1/\bar{L}$ cannot be
large.

The power-law dependence on $\bar{L}/L_1$
in Eq.~(\ref{Eq:grho}) is expected in a non-conformal effective theory,
in presence of relevant operators, in which
case there cannot be a substantial scale separation between
the UV cut-off and the mass scale $L_1$ of the effective theory itself.
This result agrees with naive dimensional analysis counting.
For instance, taking $\bar{L}=L_0$ implies that the model is strongly coupled,
unless $(L_1/\bar{L})^{n-1}\ll 4\pi$, which implies a very low cut off, and
the impossibility of describing  the resonances as weakly coupled.

Notice that this result depends smoothly on $n\gsim 1$. But in
trying to  extend the analysis to $n<1$ one immediately faces a
problem. For instance, for $L_0\rightarrow\bar{L}\ll L_1$, $n<1$,
and keeping $g^2/L$ fixed, the effective coupling becomes
vanishingly small. This behavior would imply that, in the region
of the parameters space in which the theory admits an effective approach, the original
conformal theory flows into a new phase that is described by a new
\EFT which has effectively a weaker coupling. This violates all
possible logical expectations, according to which such a phase
transition always drives the theory towards stronger coupling,
such that the new \EFT has always a smaller number of light
degrees of freedom, and hence a larger expansion parameter.
Though not rigorous, this argument seems to support the hypothesis
that only $n\gsim 1$ is an admissible choice.

From the phenomenological point of view, one way to assess how strongly
coupled is the first resonance, is to consider $\gamma_1$,
the first excited mode with the quantum numbers of a photon, and compare its
partial width into two standard model fermions  $f$
to the partial width into two on-shell $W$ bosons, namely:
\beqs
\frac{\Gamma(\gamma_1\rightarrow f\bar{f})}
{\Gamma(\gamma_1\rightarrow W^{+}W^{-})}
&\approx&
\frac{8\alpha}{3}R_1\,\frac{48\pi}{g_{\rho}^2}\,\simeq\,\frac{\pi R_1}{g_{\rho}^2}\,.
\label{Eq:width}
\eeqs
For a weakly-coupled theory this approximate estimate should be $O(1)$ or bigger.
In other words, a rough estimate of the width of the first resonance
gives $\Gamma \approx g_{\rho}^2M_1/(48\pi)$, and hence the approximation of
treating this resonance as infinitely narrow (as expected at large-$N_c$)
makes sense only as long as $g_{\rho}^2$ is at most some $O(1)$ number.
A more detailed study of this quantities, and the phenomenological
consequences relevant at LHC energies, will be discussed in a
subsequent  study.

\section{Phenomenological implications}

\subsection{Spectrum and couplings to the currents}

We start with a numerical analysis of the spectrum and couplings of the
vectorial excited states. We perform the numerical analysis because the results
discussed in the previous section for these quantities give only semi-quantitative
approximate expressions.
Since we always consider values of $m$ and $\bar{m}$ that are small compared to
$1/L_1$, the results  apply also to the axial-vector modes,
irrespectively of the choice of localizing  the
symmetry-breaking effects at $L_1$ or at $\bar{L}$.

\begin{figure*}[t]
\begin{center}
\includegraphics[width=0.8\linewidth]{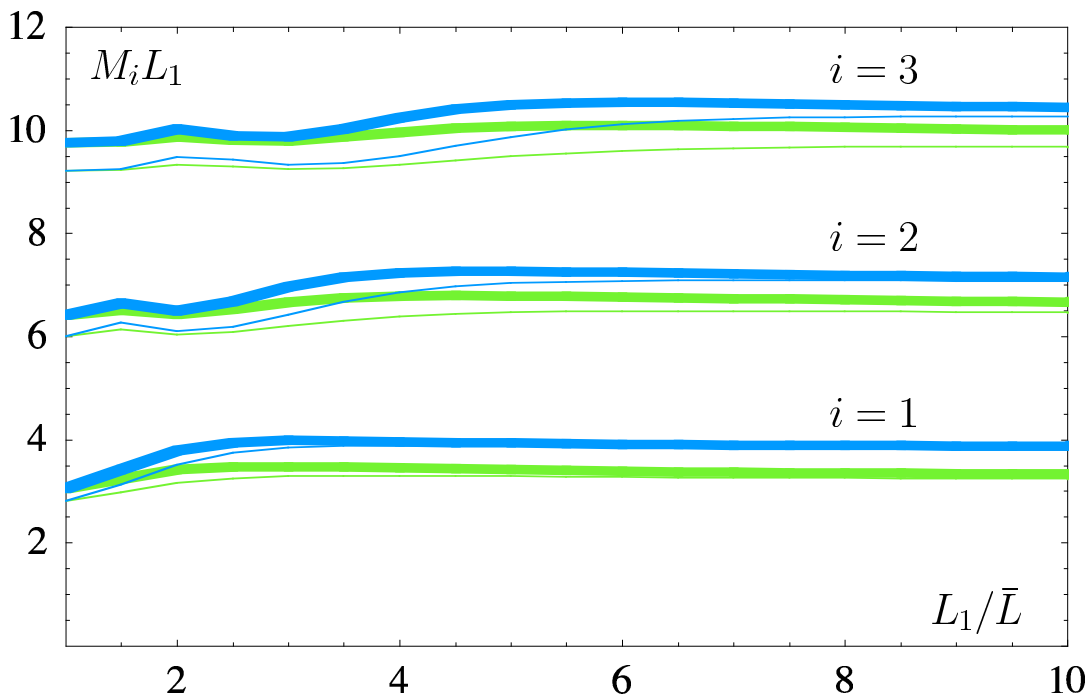}
\caption{Masses $M_iL_1$ of the first $i=1,2,3$
excited vector modes, as a function of $L_1/\bar{L}$.
The four curves are drawn for $n=2$ (green) and $n=3$ (cyan),
and for $L_0=L_1/20$ (thick line) and $L_0=L_1/100$ (thin line).
\label{Fig:masses}}
\end{center}
\end{figure*}

The masses $M_i$ depend in a complicated way on
 $L_1$, $\bar{L}$, $L_0$, and $n$.
 In Figure~\ref{Fig:masses} we plot the mass (in units of $1/L_1$)
 for the first three excited states,
 as a function of $L_1/\bar{L}$.
 We compare four choices of the relevant parameters,
 characterized by $n=2,3$ and by the choice of the UV cut off $L_0=L_1/20$
 and $L_0=L_1/100$.

As anticipated, the masses are larger than in the $n=1$ case (pure AdS),
which is recovered when $\bar{L}=L_1$. The enhancement is only moderate,
it affects the heavier states only for large values of $L_1/\bar{L}$,
and is proportional to $n$.

\begin{figure*}[t]
\begin{center}
\includegraphics[width=0.8\linewidth]{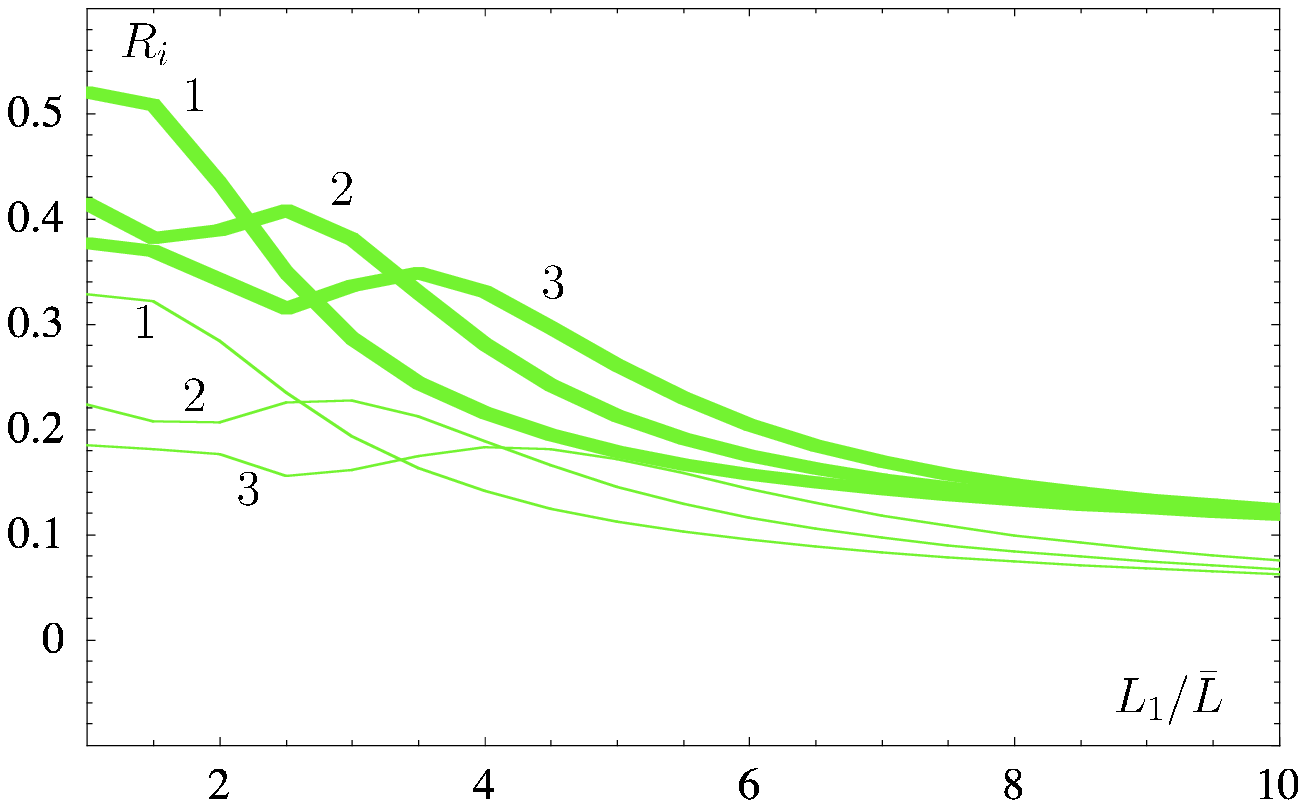}\\
\caption{Relative coupling $R_i$ to the currents of the first $i=1,2,3$
excited vector modes, as a function of $L_1/\bar{L}$.
The curves are drawn for $n=2$,
and for $L_0=L_1/20$ (thick line) and $L_0=L_1/100$ (thin line).
\label{Fig:R2}}
\end{center}
\end{figure*}

\begin{figure*}[t]
\begin{center}
\includegraphics[width=0.8\linewidth]{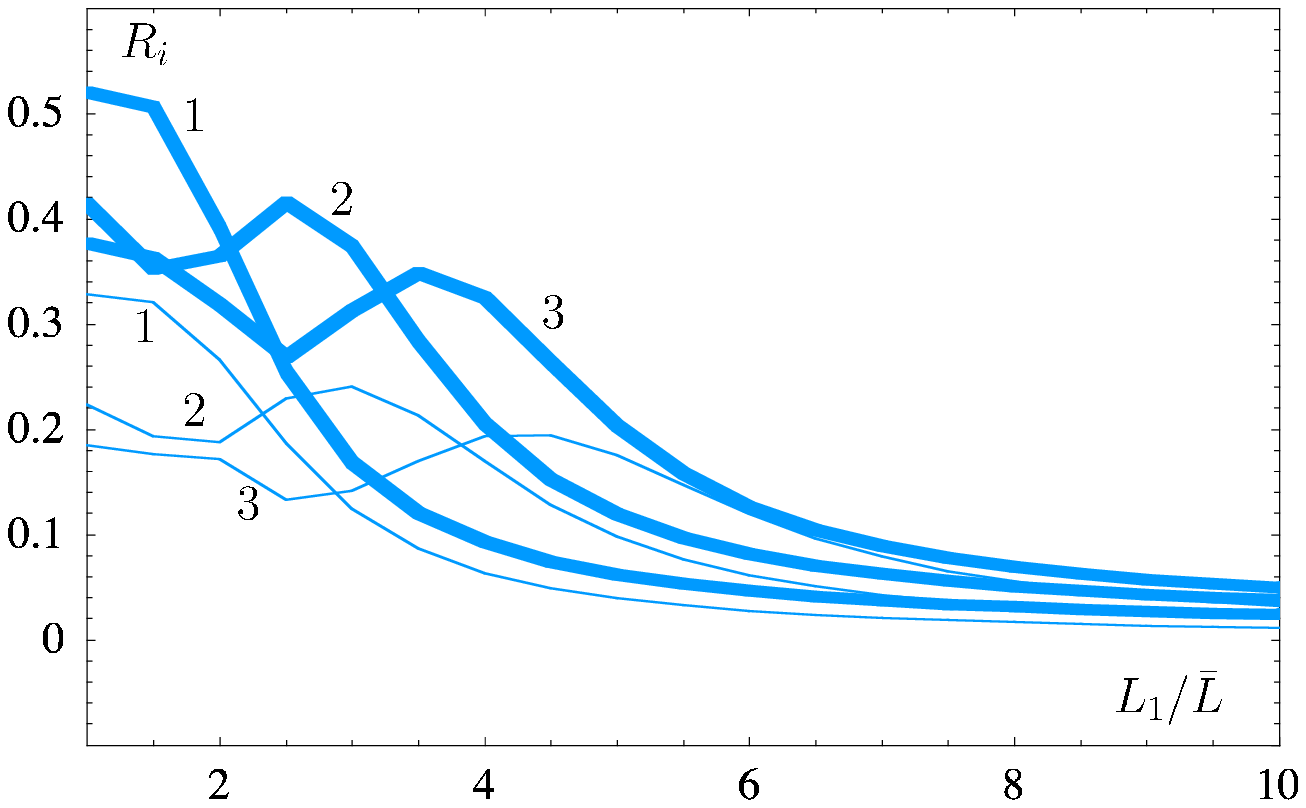}
\caption{Relative coupling $R_i$ to the currents of the first $i=1,2,3$
excited vector modes, as a function of $L_1/\bar{L}$.
The curves are drawn for  $n=3$,
and for $L_0=L_1/20$ (thick line) and $L_0=L_1/100$ (thin line).
\label{Fig:R3}}
\end{center}
\end{figure*}

The coupling $R_i$ is, in the pure AdS case, a monotonically decreasing
function of the excitation number $i$. In Figures~\ref{Fig:R2} and~\ref{Fig:R3}
 we plot the numerical results
obtained for this quantity, for the same choices of parameters
used for the masses. In going from $L_1/\bar{L}=1$ (pure AdS) to
larger values and/or to large $n$, a suppression of the coupling
is obtained for the lightest state. This suppression is a very big
effect, and it  becomes relevant at large values of $L_1/\bar{L}$.
As a result, for instance in the case $n=3$, with
$L_1/\bar{L}\gsim 4$ the third resonance has the strongest
coupling, followed by the second and by the first.

\subsection{Self-couplings and symmetry-breaking}

We want the 5D action to define a reasonable \EFT treatment of the
strong dynamics and of the resulting  \EWSB effects, with a
well-behaved perturbative expansion. We implement this requirement
by imposing the bound $g_{\rho}^2 \lsim 1/2$ (a reference value
that we fix in such a way that for the choices of parameters
discussed here the ratio of partial width estimated in
Eq.~(\ref{Eq:width}) is $\gsim 1$), where $g_{\rho}$ has been
defined in the body of the previous section. In the pure AdS case
we require that $L_1/L_0 \lsim 200$, which means that the model is
very modestly sensitive to the position of the UV cut-off and,
unless extreme choices of $L_0 \ll L_1$ are used, we can neglect
the effect of $L_0$ in driving the effective coupling strong. We
can therefore impose the bound directly on the modification due to the
new non-conformal energy regime: 
\beqs
\left(\frac{L_1}{\bar{L}}\right)^{n-1} &\lsim&
\frac{(n^2-1)}{3e^2}\,. \label{Eq:bound} \eeqs

For small values of $n \simeq 1$, the bound is not relevant,
unless very large values of $L_1/L_0$ are used. We do not discuss further this case. For $n\gsim 3$ the bound is very
restrictive, and only $L_1/\bar{L}  \sim O(1)$ is allowed. This
confirms the intuitive notion that if large power-law deviations
are allowed over a large energy window, the model is strongly
coupled and does not admit a perturbative and controllable \EFT
expansion. For $n= $ 2 -- 3, values of $L_1/\bar{L} \sim $ 3 -- 8 are
compatible with the requirement that the \EFT be weakly coupled,
and offer an interesting possibility from the phenomenological
point of view. We focus  on this possibility.

The effects of symmetry breaking are encoded in the estimate of
$\hat{S}$. This is the quantity that ultimately sets a bound on
$L_1$, and hence on the mass of the excited resonances. If the
symmetry-breaking effects are localized at $L_1$, the analytical
expression derived in Eq.~(\ref{Eq:ShatL1}) shows that, for all
practical purposes, the bounds are the same as those obtained in the pure
AdS case, $L_1 \lsim 1$ TeV$^{-1}$. This is the case because the only
sizable suppression factors are the $1/(n+1)$ and the
$\bar{L}/L_1$ terms, but at large values of $n$ only $\bar{L}/L_1
\sim 1$ is allowed.

\begin{figure*}[t]
\begin{center}
\includegraphics[width=0.8\linewidth]{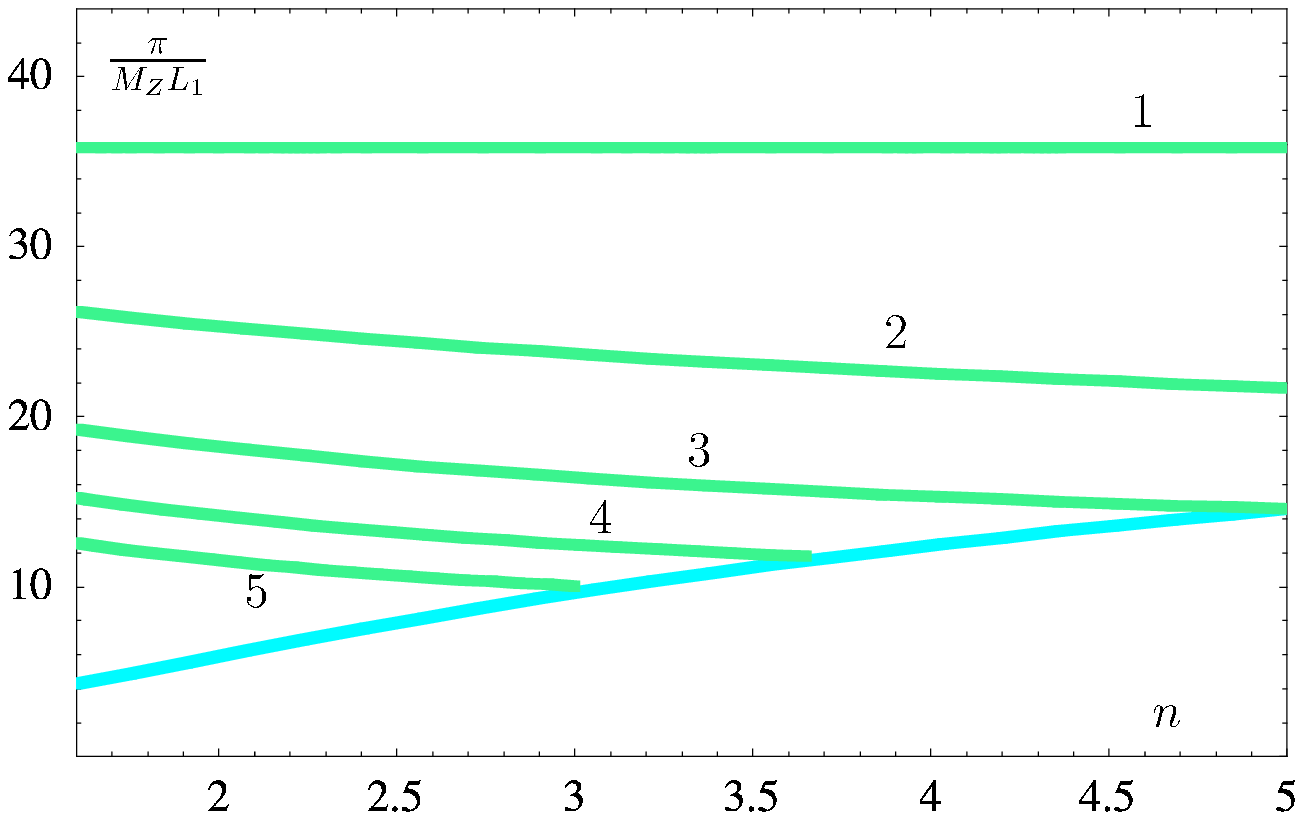}
\caption{Lower bound on $\pi/(M_Z L_1)$ as a function
of $n$ in the case in which symmetry-breaking takes place at $\bar{L}$.
The green curves are obtained using $L_1/\bar{L}=1,2,3,4,5$
The cyan curve is obtained by using the limiting value of $L_1/\bar{L}$
such that $g_{\rho}$ be perturbative.
We interrupt the (green) curves obtained at constant $L_1/\bar{L}$
at the value of $n$ for which  Eq.~(\ref{Eq:bound}) would not be satisfied,
which is at the intersection with the cyan curve.
\label{Fig:Y}}
\end{center}
\end{figure*}

Let us discuss  the case in which
symmetry-breaking takes place at $\bar{L}$. In order to assess how
sizable the reduction in the experimental bounds is, we require
that $\hat{S}<0.003$, and calculate the minimum  value of
$1/L_1$ which is compatible with this bound, using the
expression in Eq.~(\ref{Eq:ShatLb}). We show the result in
Figure~\ref{Fig:Y} assuming various values of
$L_1/\bar{L}$. We plot, as a function of $n$, the lower bound
for $\pi/(M_ZL_1)$ -- which, up to  boundary effects and model-dependent shifts, gives a reasonable estimate of the ratio $M_1/M_Z$ (see
Figure~\ref{Fig:masses}) -- starting from the pure AdS case, but without exceeding the ($n$-dependent) bound
in Eq.~(\ref{Eq:bound}).

In the pure AdS case ($L_1/\bar{L}=1$),
the lower bound in Figure~\ref{Fig:Y} implies (using the
experimental value of $M_Z$)  $M_1\gsim 3$ TeV, and $M_2 \gsim$ 6-7 TeV. Going to larger values of $L_1/\bar{L}$ allows for a very
significant reduction of such bounds, even when this ratio is
small enough to be compatible with the requirement that the
effective coupling $g_{\rho}^2$ be smaller than $1/2$. As a result, the value of
the scale  $1/L_1$ can be greatly reduced. Values such as $M_1\sim
1.5$ TeV, $M_2 \sim 3$ TeV and $M_3 \sim 4.5$ TeV are not excluded
experimentally.

A detailed calculation of the coupling to the currents and of the partial widths
is necessary in order  to draw firm quantitative conclusions,
but these preliminary estimates indicate that the first three resonances
have  $R_i \sim 0.15 - 0.35$, while $g_{\rho}^{(i)\,2} \lsim 0.5$.
These resonances should have a sizable branching fraction in
standard-model fermions, and a sizable production cross-section
in Drell-Yan processes. In particular, for this range of masses and couplings,
LHC has a  good chance of detecting all of these states
even at moderate integrated luminosity,
 by combining data on $\mu^{+}\mu^{-}$ and $e^{+}e^{-}$
final states.

\section{Discussion}

The starting point for the construction of an \EFT
description of dynamical \EWSB is the assumption that some
fundamental, possibly asymptotically free, field theory, defined
in the far UV, flows towards an (approximate) strongly-coupled
fixed-point in the IR. Accordingly, there is  a regime at
intermediate-to-low energies in which the (walking) theory can be
described by a weakly-coupled five-dimensional model, in the
spirit of the AdS/CFT correspondence. The presence of a deformation away from the AdS metric---in the form of some operator that  becomes relevant
and dominates the dynamics at long distances---drives the model
away from the fixed point (inducing the loss of conformal
behavior), produces non-trivial condensates (which trigger
spontaneous electro-weak symmetry breaking), and ultimately
leads the theory towards confinement (and hence introducing a
mass gap in the spectrum of bound states).

This paper proposes a toy-model that allows for a quantitative
study of the effects that such a relevant deformation might have
on the low-energy observable quantities, in the regime at and below
the LHC relevant energies. The basic idea is to parameterize the
effects of such a deformation in terms of a power-law departure
from the AdS background over a limited energy window just above
the scale of confinement. This treatment proves to be useful thanks to its intrinsic simplicity and the lack of any more systematic (calculable) approach. It
has its limitations as well. Hence we summarize and critically analyze
our results, in order to draw some important model-independent
conclusion and in order to highlight the areas where more work,
and possibly some guidance from the experimental data to come, are
necessary.

First of all, the type of modification of the background we propose has
a very modest effect on the spectrum of composite resonances.
The properties of such spectrum are still determined by the presence of a hard-wall
in the IR, that acts both as a regulator and as a physical scale determining the
mass gaps and spacings.
 It is inappropriate to believe that this model can describe
accurately  more than a handful of resonances, and one should be
very careful when talking about resonances with large excitation
number $i$. Yet, the model-independent message here is quite clear,
and very important. While the spectrum is substantially
independent of the possible presence, and type, of deformation
that is driving the theory away from the fixed point in the IR,
the effective couplings of the resonances, both to other
resonances and to the \SM fermions, are very sensible to the
departure from conformality that this deformation is introducing.

The calculation of the coupling to the currents and the estimate
of the self-couplings show a large departure from the
expectations based on the pure AdS case, in presence of the same
regulators in the IR and in the UV. The coupling to the currents
is suppressed, and the suppression in not a universal effect, but
rather it is different for different resonances. The self
couplings are enhanced with respect to the pure AdS case, following
the four-dimensional intuition. This poses some important
limitation on how long it is admissible to assume  that it will
take for the theory to flow from the region in proximity
of the IR fixed point, where it is walking, to the new phase
transition at which confinement takes place. It is very
encouraging that our estimates indicate that this regime, though
limited, might be long enough to allow for very sizable $O$(2-4)
effects to result, without spoiling the calculability of the \EFT
that the AdS/CFT language is supposed to provide.

The deformation responsible for the loss of conformal symmetry
might or might not be related with electro-weak symmetry breaking.
If not, then \EWSB is triggered at the same scale as
confinement, as is the case for QCD. In this case this model
allows us to say that we do not expect any significant
modification of the precision electro-weak parameters and of the
coefficients of the electro-weak chiral Lagrangian with respect to
the results obtained in the pure AdS background. In this case, the
couplings of the excited states are the only observable quantities
carrying information about the existence of an energy regime above
the scale of confinement where the dynamics is not conformal.

At large-$N_c$ or in presence of a complicated
fermionic field content in the fundamental theory, the chiral
symmetry breaking condensates may form at a temperature
 larger than the scale of confinement. In this case,
the formation of such condensates might itself be the deformation
that drives the theory away from the fixed point, and that leads
to confinement at some lower scale. The phenomenological
consequences of such a scenario are relevant not only for the LHC,
but even in analyzing LEP and TeVatron data. Our simple model
allows us to show that it is  reasonable to expect that
in this case the estimates of the coefficients of the chiral
Lagrangian (we focused on $\hat{S}$ because best known and most
model-independent) might be suppressed by large numerical factors, without entering a strongly coupled regime for the
effective field theory, and with a resulting drastic reduction of
the experimental bounds on the masses of the lightest new spin-1
states (techni-$\rho$). This toy-model highlights the fact that,
whatever the fundamental theory is in the far UV,  if the
dynamics contains a mechanism leading to a separation of the
scales of chiral symmetry breaking and confinement, then the
expectations for $\hat{S}$, and for other precision parameters
related with isospin breaking, can be changed drastically. At the
LHC, this implies that, without requiring any additional custodial
symmetry, nor any fine-tuning, the dynamics itself might be
compatible with the detection of the first  two or even three
excited states, which would provide unmistakable evidence for a
strongly-coupled origin of electro-weak symmetry breaking.

The techniques used here, and the choices of parameters we make,
are  affected by systematic uncertainties. The
numerical results we obtain are to be taken as an indication of
what is possible, rather than as robust predictions. Yet, part of
the results are completely general: for any admissible choice of
$L_1/\bar{L}$, of $n > 1$ and of the position in the fifth
dimension at which we localize the symmetry-breaking terms, there
is always a suppression of the coupling of the vector
mesons to the currents, an enhancement of their self-couplings,
and a suppression of $\hat{S}$. These are quantitative
model-independent results, indicating that for these quantities
the pure AdS case yields always a limiting, conservative estimate.
And they all point in the direction of making the experimental
searches at the LHC easier.

\vspace{1.0cm}
\begin{acknowledgments}
The work of MP is supported in part  by the Department of Energy
under the grant DE-FG02-96ER40956, and  by the Wales Institute of
Mathematical and Computational Sciences. The work of MF and LV is
partially supported by MIUR and the RTN European Program
MRTN-CT-2004-503369. MF thanks SISSA for the kind hospitality.

\end{acknowledgments}


\end{document}